\pdfoutput=1
\documentclass[sigconf,screen]{acmart}
\usepackage{xspace}
\usepackage{multirow}
\usepackage{enumitem}
\usepackage{algorithm2e}
\usepackage{algorithmicx}
\usepackage{algpseudocode}
\usepackage{threeparttable}
\usepackage{booktabs}
\usepackage{svg}
\usepackage[most,skins]{tcolorbox}
\usepackage{xcolor}      % Required for custom colors
\usepackage{makecell} % Add this in the preamble
\usepackage{xcolor}
\usepackage{amsmath}     % Required for math formatting
\definecolor{DarkGreen}{RGB}{0,100,0} % RGB values for dark green

\newcommand{\tool}[0]{CodeThinker\xspace}

\newcommand{\dataset}[0]{SVRC\xspace}

\AtBeginDocument{%
  }

\usepackage{cleveref}

\setcopyright{acmlicensed}
\copyrightyear{2025}
\acmYear{2025}
\begin{document}
\setcopyright{none} % to remove the copyright notice
\settopmatter{printacmref=false} % to remove the ACM Reference Format
\renewcommand\footnotetextcopyrightpermission[1]{}

% \title{Harnessing Community Knowledge for Reasoning Code Generation: An Adaptive and Structured Code Reasoning Framework}
\title{Think Like Human Developers: Harnessing Community Knowledge for Structured Code Reasoning}


\author{Chengran Yang}
\affiliation{%
  \institution{Singapore Management University}
  \country{Singapore}}
\email{cryang@smu.edu.sg}

\author{Zhensu Sun}
\affiliation{%
  \institution{Singapore Management University}
  \country{Singapore}}
\email{zssun@smu.edu.sg}

\author{Hong Jin Kang}
\affiliation{%
  \institution{University of Sydney}
  \country{Australia}}
\email{hongjin.kang@sydney.edu.au}

\author{Jieke Shi}
\affiliation{%
  \institution{Singapore Management University}
  \country{Singapore}}
\email{jiekeshi@smu.edu.sg}

\author{David Lo}
\affiliation{%
  \institution{Singapore Management University}
  \country{Singapore}}
\email{davidlo@smu.edu.sg}
% \affiliation{%

\renewcommand{\shortauthors}{Chengran et al.}


\begin{abstract}
Large Language Models (LLMs) have significantly advanced automated code generation, yet they struggle with complex coding tasks requiring multi-step logical reasoning. High-quality reasoning data is crucial for improving LLMs' reasoning capabilities, but such datasets remain scarce. Existing approaches either rely on computationally expensive reinforcement learning (RL) or error-prone reasoning chains synthesized by LLMs, posing challenges in scalability and accuracy.

To address this challenge, we propose SVRC (Structured and Validated Reasoning Chains for Code Generation), a novel framework that mines, restructures, and enriches reasoning chains from community-driven discussions on software engineering platforms. 
SVRC refines unstructured and incomplete discussions of coding problems by aligning them with Software Development Life Cycle (SDLC) principles, ensuring that reasoning chains capture real-world problem-solving strategies and support iterative refinement.

To evaluate the effectiveness of SVRC, we introduce CodeThinker, an LLM fine-tuned on 12,444 reasoning-augmented samples generated by SVRC. Experiments on LiveCodeBench show that CodeThinker surpasses its base model by 42.86\% on medium-level code problems in terms of pass@1 and outperforms GPT-4o-mini and GPT-4o by 73.14\% and 115.86\%, respectively. Our ablation study further highlights that each component of SVRC contributes to the reasoning capabilities of CodeThinker.

\end{abstract}

\maketitle

\section{Introduction}
Large Language Models (LLMs) have revolutionized software engineering with exceptional performance across a spectrum of coding tasks~\cite{sun2024ai,hou2023large,fan2023large,wang2024software}, ranging from code summarization ~\cite{ahmed2022few,ahmed2024automatic,10.1145/3551349.3560421}, vulnerability detection~\cite{zhou2024large,zhang2025benchmarking}, to program repair~\cite{jiang2023impact,jin2023inferfix,nguyen2023multi}. Among these, code generation~\cite{jiang2024survey,rasheed2025large, yang2024acecode}—the process of generating code from natural language requirements—has been one of the most impactful applications of LLMs, enabling automatic tools like GitHub Copilot~\cite{githubGitHubCopilot} and Cursor Editor~\cite{cursorCursorCode} to assist developers. 
However, despite their ability to generate syntactically correct code, recent studies indicate that LLMs often struggle to fully understand the natural language task descriptions and fall short in constructing solutions that require advanced logical reasoning~\cite{wang2024large, dou2024s}, highlighting an urgent need to improve their code reasoning capabilities.

Being data-driven, improving LLMs' capacity for code reasoning typically requires training with high-quality datasets that explicitly encode reasoning steps for complex coding problems. 
However, manually creating such datasets is time-consuming and costly~\cite{han2022folio0, buse2011benefits, he2025code}, demanding advanced programming expertise, which has resulted in a scarcity of large-scale code reasoning datasets.
To address this, researchers have explored alternative training methods that leverage synthetic reasoning chains generated by LLMs. For example, OpenAI's o1~\cite{jaech2024openai} and DeepSeek's R1~\cite{guo2025deepseek} utilize reward optimization to encourage LLMs to autonomously produce step-by-step reasoning. 
However, their reliance on large-scale base models and associated high computational costs significantly restrict their scalability and broader adoption.
% Other approaches involve using reasoning traces generated by larger LLMs to fine-tune smaller models~\cite{toshniwal2024openmathinstruct,tang2024mathscale}.
Other approaches involve fine-tuning LLMs using reasoning traces from math and science problems generated by larger LLMs~\cite{toshniwal2024openmathinstruct,tang2024mathscale, muennighoff2025s1}.
Yet, these synthetic reasoning chains can be error-prone since they heavily depend on the intrinsic reasoning abilities of LLMs, which may not always generate reliable reasoning steps comparable to those produced by humans.
This becomes especially problematic when flawed reasoning still results in a correct final answer~\cite{xia2024evaluating}, as it can mislead the fine-tuning process.
Thus, there is a pressing need for a scalable as well as reliable method to construct code-generation datasets enriched with accurate reasoning chains.

\begin{figure}[!t]
    \centering
    \includegraphics[width=0.37\textwidth]{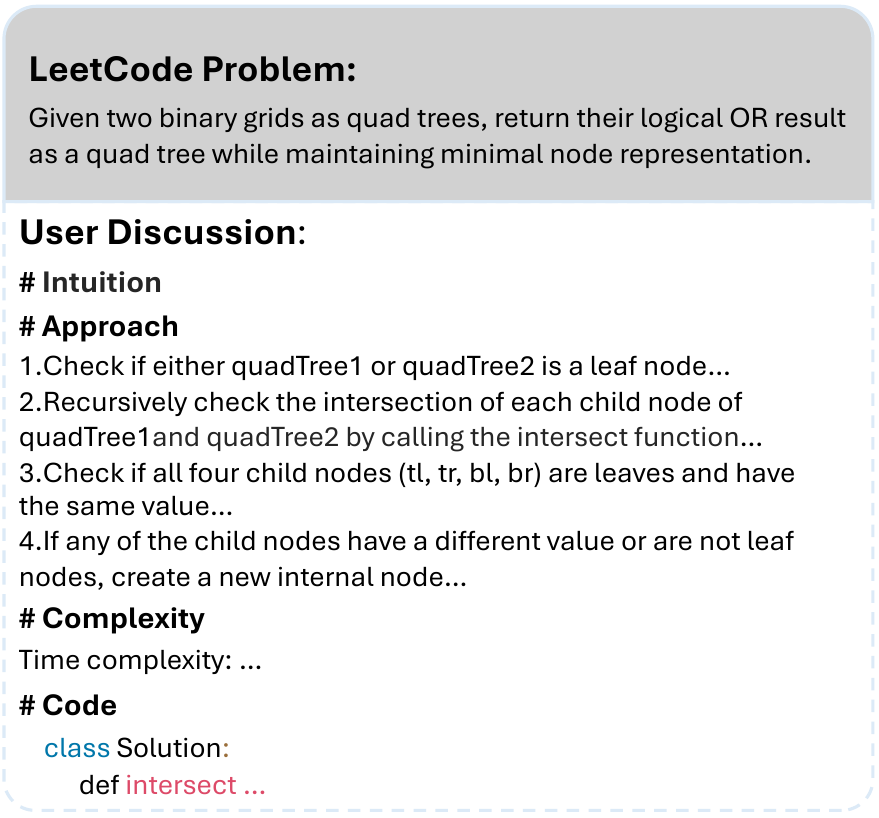}
    \caption{An example of community discussions on LeetCode.}
    \label{fig:discussion-example}
    \vspace{-3mm}
\end{figure}

To address this need, we turn to software engineering community platforms such as LeetCode~\cite{leetcode} and Stack Overflow~\cite{stackoverflow}, which host extensive collections of coding problems, corresponding solutions, and detailed discussions. As illustrated in~\Cref{fig:discussion-example}, these discussions offer authentic, peer-reviewed insights into developers' reasoning and problem-solving strategies, making them a valuable yet underutilized resource for mining human reasoning patterns. Such reasoning patterns hold great potential to enhance LLMs' capabilities for human-like code reasoning.

However, harnessing community discussions on coding problems to enhance LLMs' code reasoning ability is challenging because they are often unstructured, scattered, and incomplete. 
% Directly incorporating these disorganized discussions into training datasets can prevent models from effectively understanding and learning reasoning patterns. 
We show that directly applying disorganized community discussions to fine-tune LLMs prevents models from effectively understanding and learning reasoning patterns (as discussed in Section~\ref{sec:ablation1}).
Moreover, human developers rarely arrive at a perfect solution on their first attempt; instead, they follow an iterative refinement process—reconsidering requirements, debugging, and optimizing their code in cycles. 
However, most community discussions focus only on the final correct solution, omitting the trial-and-error reasoning that led to it, which is crucial for LLMs to capture and mimic the iterative problem-solving process. This motivates us to develop a method that not only extracts and structures reasoning chains from successful solutions but also enriches them with the iterative refinement process that reflects real-world coding practices.

In this paper, we propose \dataset{} (Structured and Validated Reasoning Chains for Code Generation), a novel method for extracting and augmenting code generation datasets with structured reasoning chains mined from community platforms. SVRC transforms unstructured problem-solving insights into well-formed, iterative reasoning steps aligned with Software Development Life Cycle (SDLC) principles~\cite{10.1145/1764810.1764814}—a systematic software development process encompassing phases such as problem revisiting, planning, design, coding, testing, and refinement. Specifically, \dataset{} begins by collecting coding problems along with their discussions, applying heuristic rules to retain only high-quality content. These discussions are then processed to fill in missing sections of reasoning chains using state-of-the-art LLMs while also perturbing code formats to enhance dataset diversity and ensure adherence to a structured reasoning format. At this stage, the extracted reasoning chains provide coherent steps for a code solution but fail to capture the iterative refinement process of human developers. To address this, we draw inspiration from SDLC principles and employ an LLM to refine the raw reasoning chains. This refinement process mirrors key SDLC stages, including requirement analysis, planning, solution design, and solution development with testing. 
Additionally, we expand the reasoning chains by explicitly incorporating iterative solution refinement, enabling them to systematically identify issues, analyze their root causes, and progressively optimize solutions.
The resulting structured reasoning chains are comprehensive and well-organized, making them suitable for fine-tuning LLMs to emulate human-like reasoning in code-generation tasks.

To evaluate the effectiveness of \dataset{}, we fine-tune Qwen2.5-Instruct-32B and release \tool{}, an LLM with enhanced code reasoning capabilities, as a proof of concept. The fine-tuning process utilizes 12,444 data samples synthesized from \dataset{} and sourced from LeetCode discussions.
We evaluate the performance of \tool{} on LiveCodeBench, which continuously aggregates the latest coding problems from diverse competitive programming platforms such as LeetCode, Codeforces, and AtCoder. To ensure a fair evaluation, we exclude all LeetCode problems used during fine-tuning.
The experimental results show that, by fine-tuning with \textbf{only 12,444} data samples, \tool{} enables advanced code reasoning and outperforms its base model on medium-level problems by 42.86\% in terms of pass@1 and achieves on-par performance on easy-level problems. 
Furthermore, \tool{} achieves better performance than GPT-4o-mini and GPT-4o in addressing medium-level code problems by 73.14\% and 115.86\% in terms of pass@1, respectively.
Our ablation study further illustrates that each stage of \dataset{} contributes to the final performance of \tool{}.

In summary, our contributions are as follows:
\begin{itemize}[leftmargin=*]
    \item We introduce \dataset{}, a novel pipeline that transforms unstructured and incomplete community discussions into structured and diverse reasoning chains with iterative refinements.
    \item We release SVRC$_{LC}$, the first reasoning-augmented code generation dataset that is synthesized from high-quality human discussions on LeetCode. 
    \item We propose \tool{}\footnote{https://huggingface.co/Chengran98/codethinker}, a reasoning-augmented LLM fine-tuned on SVRC$_{LC}$, which outperforms existing models on challenging code problems. More model checkpoints will be available soon.
    \item We conduct thorough ablation studies to illustrate the effectiveness of each stage of \dataset{}. 
\end{itemize}

\section{Motivation}
\subsection{Task Formulation}
This paper focuses on reasoning-augmented code generation, where a code generator explicitly generates reasoning steps before producing the final solution for a given problem statement.
Formally, given a problem statement $\mathit{P}$, the goal of reasoning-augmented code generation is to generate a sequence of reasoning steps $\mathit{R} = \{r_1, r_2, \cdots, r_n\}$ in natural language and a final code $\mathit{C}$ that is derived from $\mathit{R}$ and solves $\mathit{P}$.

\subsection{How Developers Reason in Discussions}\label{sec:motivation}
We show one example of discussions on LeetCode in \Cref{fig:discussion-example}, in which the user provides a solution~\cite{leetcodeexamplediscussion} to a coding problem~\cite{leetcodeexamplequestion} and corresponding insights towards the solution.
Specifically, the user breaks down and explains his solution into a chain of steps: applying conditional checks to handle leaf nodes, recursively merging child nodes, and simplifying the tree by merging identical children.
This chain of steps infers a reasoning path that leads to the final solution and can be used to guide LLMs to reason.

However, such discussions still suffer from quality and diversity issues.
For example, the discussion in \Cref{fig:discussion-example} leaves the Intuition section empty, making it difficult for an LLM to fully grasp the underlying problem-solving rationale.
Besides, all Python solutions on LeetCode are written in a fixed format, i.e., a class definition named \textsc{Solution}.
Fine-tuning LLMs with such discussions may raise the risk of LLM overfitting to a fixed format.
We observe that fine-tuning LLMs with raw discussions from one platform significantly hinders their generalization ability on code problems from other platforms in \Cref{sec:ablation1}.
The above observations motivate us to design a pipeline that transforms and enriches raw discussions of code problems into structured and diverse reasoning chains.

\subsection{Related Work}
We introduce two lines of related works that focus on reasoning-augmented code generation. 
Traditional approaches to reasoning-augmented code generation primarily leverage the intrinsic reasoning capabilities of LLMs by prompting them to generate intermediate reasoning steps during the inference stage. Techniques such as Chain-of-Thought (CoT) prompting~\cite{wei2022chain} and its variants like Tree-of-Thought~\cite{yao2023tree} and Graph-of-Thought~\cite{besta2024graph} have been proposed to enhance LLMs' reasoning capabilities.
However, those approaches heavily rely on LLM's inherent but often limited reasoning ability~\cite{wei2022chain, yao2023tree, besta2024graph}, while code reasoning is still a challenge for many LLMs~\cite{wang2024large, dou2024s}.
Different from the above approaches, we focus on fine-tuning LLMs with reasoning patterns mined and enriched from community discussions.

Recently, researchers have proposed fine-tuning methods to enable the reasoning ability of LLMs. 
Some works fine-tune LLMs with extensive code corpora to improve their code understanding and reasoning abilities, e.g., DeepSeekCoder~\cite{guo2024deepseek}, Qwen2.5-Coder~\cite{hui2024qwen2}, and CodeLlama~\cite{guo2024deepseek}.
However, those LLMs excel at handling syntactic complexity but often lack the reasoning abilities of code problem-solving.
RL-based LLMs, e.g., OpenAI's o1~\cite{jaech2024openai} and DeepSeek's R1~\cite{guo2025deepseek}, rely on reward optimization to train LLMs toward generating step-by-step reasoning during the inference stage. However, their reliance on large base models and extensive computational resources constrains their broader adoption. 
Researchers also fine-tune LLMs with reasoning traces of math or science problems~\cite{toshniwal2024openmathinstruct,tang2024mathscale, muennighoff2025s1} synthesized from larger LLMs to improve their math reasoning abilities.
However, the synthetic reasoning chains can be unreliable, potentially misleading the fine-tuning process~\cite{xia2024evaluating}.
Different from the aforementioned approaches, we focus on fine-tuning LLMs with ground-truth reasoning patterns mined from community discussions.
Such reasoning chains are authentic and validated (i.e., effective), eliminate the need for costly manual annotations (i.e., efficient), and leverage continuously expanding community data repositories (i.e., scalable). Moreover, by refining these reasoning chains according to established software engineering best practices, our approach specifically tailors the fine-tuning process toward code reasoning tasks.

\section{SVRC}
\label{sec:svrc}
\begin{figure*}
    \centering
    \includegraphics[width=0.80\textwidth]{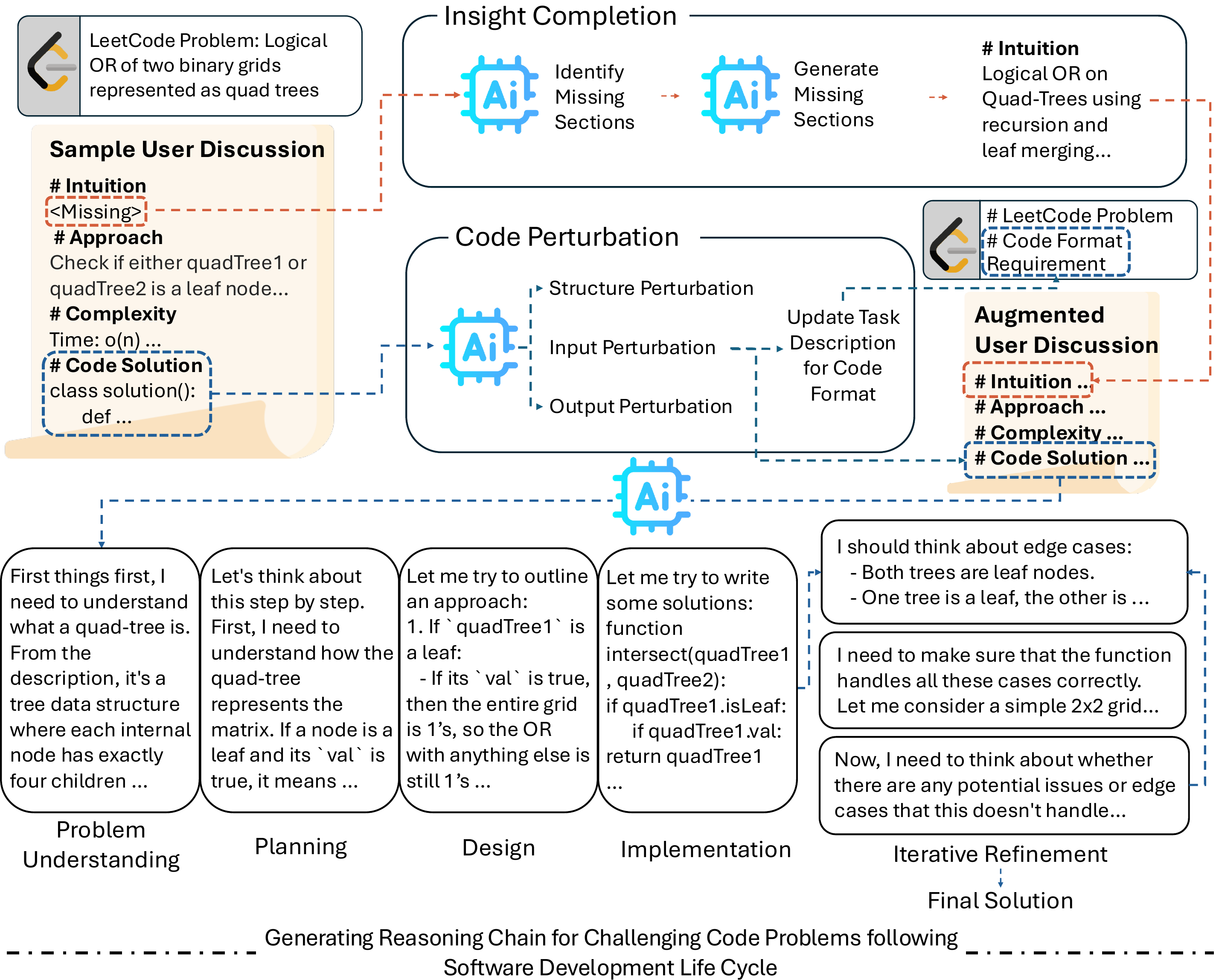}
    \caption{The pipeline of \dataset{}. \dataset{} begins with extracting community discussions and corresponding code solutions from community platforms. It then identifies and completes missing sections in discussions, and introduces format perturbations to code solutions to enhance original discussions. Finally, \dataset{} enriches the reasoning chains by rewriting the original discussion guided by software development principles and introducing refinement loops. }
    \label{fig:overall_pipline}
\end{figure*}

In this section, we describe the pipeline of \dataset{}. 
\dataset{} aims to transform and enrich community-contributed human discussions for code problems into structured, diverse, and validated reasoning chains.
Each output of \dataset{} includes a triple $(Q, A, R)$, in which $Q$ is a programming question, $A$ is the corresponding ground-truth code solution, and $R$ is a reasoning chain outlining the problem-solving reasoning steps leading to the solution.
We show the pipeline of crafting reasoning chain $R$ in Figure~\ref{fig:overall_pipline}, where each reasoning chain $R$ is systematically constructed through a two-stage pipeline:
\begin{itemize}[leftmargin=*]
    \item \textbf{Community Knowledge Extraction}: Collect and filter problem-solving insights and corresponding ground-truth code solutions from high-quality discussions of coding problems on community platforms. Then complete missing insights of discussion and introduce format perturbations to code solutions.
    \item \textbf{Reasoning Chain Enrichment}: Transform discussions into structured reasoning chains guided by Software Development Life Cycle (SDLC) principles and enhanced through iterative refinements.
\end{itemize}

We emphasize that the design of \dataset{} is programming language agnostic and minimally dependent on specific platforms.
Its only dependency is a voting mechanism within community discussions, which serves as a proxy for discussion quality, allowing the automatic selection of high-quality discussions without manual curation.
Such voting-based discussion forums are widely used across various coding platforms, including blogs on CodeForces, forums on CodeChef, SPOJ, LeetCode, and widely-used community Q\&A platforms like Stack Overflow.

\subsection{Community Knowledge Extraction}
The first step in \dataset{} involves extracting problem statements and high-quality problem-solving discussions.
We begin by crawling freely available coding problems on the target community platform.

To gather high-quality problem-solving discussions, we leverage the number of community upvotes as a natural metric for evaluating discussion quality. 
Following existing approaches on community knowledge summarization~\cite{yang2022answer}, we extract the top 10 highest-voted discussions tagged with the target programming language for each coding problem.
Many discussions may lack textual insights, providing only code snippets without explanations.
To ensure discussions contain textual insights, we discard discussions with fewer than 100 characters of textual content.

\subsubsection{Discussion Completion}
As mentioned in~\Cref{sec:motivation}, discussions in the community are often incomplete.
To fix this issue, we employ a two-stage completion method using LLMs to infer and complete missing sections.

Referring to the typical structures of the community platform LeetCode, we identify the essential components of a comprehensive discussion.
Specifically, each discussion should include:
\begin{itemize}[leftmargin=*]
    \item \textbf{Intuition}: A high-level explanation or an initial perspective on the problem-solving strategy.
    \item \textbf{Approach}: A step-by-step breakdown of the problem-solving methodology, explicitly describing each step in a logically structured and detailed manner.
    \item \textbf{Complexity}: A precise statement of the time and space complexity of the proposed approach.
\end{itemize}

In the first stage, we employ an LLM to classify whether a given discussion misses any of the components.
Specifically, for each section of the discussion, the LLM is prompted to understand its text content and assign a binary classification label indicating whether the corresponding information is missing.
In the second stage, for sections identified as incomplete, we employ LLM to generate the missing sections, following the aforementioned format criteria and the discussion context (i.e., the problem description, existing discussion, and code solution).
The rationale behind this two-stage completion is grounded in the principle of modular reasoning, which improves control and predictability. By decoupling the identification of missing components from the generation of missing content, this approach ensures that inferred details are introduced without unintended modifications to the existing discussion by LLMs, such as altering its original discussion. 

\begin{table}[h]
    \centering
    \caption{Format Variations in \dataset{}}
        \vspace{-4mm}
    \renewcommand{\arraystretch}{1.2}
    \resizebox{0.41\textwidth}{!}{ % Resizing to fit text width
    \begin{tabular}{ll}
        \toprule
        \textbf{Category} & \textbf{Variation Type} \\
        \midrule
        \multirow{6}{*}{\textbf{Input Format}} 
        & Direct input (e.g., variables, constants) \\
        & \makecell[l]{Structured input \\ (e.g., list, tuple, dictionary)} \\
        & Batch input \\
        & Interactive input() \\
        & Boolean input  \\
        & Command-Line Arguments \\
        \midrule
        \multirow{5}{*}{\textbf{Output Format}} 
        & \makecell[l]{Structured input \\ (e.g., list, tuple, dictionary)} \\
        & Boolean output \\
        & Batch output \\
        & Print statement \\
        & Logging Output \\
        \midrule
        \multirow{3}{*}{\textbf{Structural Variations}} 
        & Class-based \\
        & Function-based \\
        & Script-based \\
        \bottomrule
    \end{tabular}
    \label{tab:format_variations}
    }
    \vspace{-4mm}
\end{table}

\subsubsection{Format Perturbation}
We observe that code solutions on coding contest platforms often adhere to a standardized template.
For example, LeetCode requires all Python3 solutions in the form of a class with a fixed class name, \textsc{Solution}, to facilitate the testing process. 
Such standardization reduces dataset diversity and increases the risk of model overfitting, limiting the model's ability to generalize across varied coding styles and programming scenarios.
Simply merging discussions from multiple platforms does not fully resolve this issue, as models continue to learn from a constrained set of template structures, restricting their adaptability.

A straightforward approach to improve code diversity is to use LLMs to rewrite solutions. However, unrestricted modifications by LLMs may introduce unintended changes, causing misalignment between the code solution and corresponding reasoning steps in discussions. 
For instance, altering variable names in code solutions could create inconsistencies if the discussion references the original names, potentially causing confusion.

To mitigate this issue, we introduce format perturbations that systematically modify the input, output, and structural formats of code solutions while preserving their functionality. Specifically, for each code solution, we generate a set of structural, input, and output format variations using predefined transformation operators (as outlined in~\Cref{tab:format_variations}). For instance, structural variations may convert a class-based solution into a function-based one, input formats may shift from direct variable assignments to standard user input mechanisms, and output formats may be adjusted to use standard printing functions.
Then, given a code solution and a set of format variations, we employ an LLM to rewrite the code accordingly while ensuring minimal functional modifications.
We ensure that LLMs operate within a predefined scope by applying controlled perturbations, preserving alignment between the code and discussion while maintaining logical coherence.

After applying these perturbations, we automatically add the code format requirement within the task description to align with the perturbed code solution. 
Specifically, based on the variations applied to the code, we prompt the LLM to generate a corresponding textual requirement for the format of the code solution, which explicitly defines the expected format of code structure, input, and output.
This requirement is appended to the original problem statement as a complementary task description, thereby ensuring consistency between the task description and the perturbed code.

\subsection{Reasoning Chain Enrichment}
\label{sec: reasonenrich}
The completed and perturbed discussion can be regarded as a raw reasoning chain with three core steps, i.e., inferring intuition, planning the approach, and estimating the complexity.
% At this stage, augmented discussions offer a structured and diverse reasoning chains of the code solution.
However, many discussions contain only one reasoning path to the solution, which makes it fail to express the cognitive process involved in iterative human problem-solving—such as evaluating alternative approaches, debugging mistakes, and refining solutions iteratively.
As a result, LLMs trained purely on a single reasoning path lack the capabilities of reasoning in depth, self-correction, and iterative refinement. 
% This limitation prevents LLMs from reasoning in-depth and refining their solutions.

To address this challenge, we take inspiration from Software Development Life Cycle (SDLC) principles~\cite{ruparelia2010software} and enrich reasoning chains accordingly. 
SDLC is a structured process that guides software development through sequential phases including requirement analysis, planning, design, implementation, testing, deployment, and maintenance.
It naturally fosters a structured and systematic thinking paradigm that aligns with software engineering practices and models how human developers analyze, design, and refine solutions in an iterative manner.
With this paradigm, we use LLMs to rewrite and enrich the existing reason chain.
By enriching them with SDLC principles, the reasoning chain can approach complex coding tasks as human developers would, by treating them as dynamic software development processes rather than rigid, procedural executions.

Guided by the SDLC principles, we define the key phases of solving coding problems as follows:
\begin{itemize}[leftmargin=*]
    \item \textbf{Problem Understanding}: Begin by thoroughly reading and analyzing the task description, clarifying requirements and constraints.
    \item \textbf{Planning}: Devise a solution strategy by breaking down the task into smaller, manageable subproblems. Compare potential approaches and select the most efficient one based on constraints.
    \item \textbf{Design}: Outline the steps needed to implement the chosen approach. This could include pseudocode, notes on algorithms, or identifying edge cases.
    \item \textbf{Implementation}: Write clean, readable, and well-documented code. Use meaningful variable names and adhere to best practices.
    \item \textbf{Testing}: Test the code with a variety of cases, including edge cases, to ensure correctness and robustness.
    \item \textbf{Iterative Refinement}: Iterative refinement begins with a sub-optimal solution that contains a common developer mistake, leading to a failure. The reasoning process should systematically identify the issue during testing, analyze its root cause, and iteratively refine the solution to optimize the code.
\end{itemize}

Notably, during the testing phase, LLMs do not interact with an execution environment but instead simulate the thinking process guided by testing, systematically anticipating potential failure cases, handling unexpected inputs, and reasoning about possible errors without actual code execution.

These key phases are translated into a prompt.
Using it, we employ an LLM to rewrite the reasoning steps of the discussions in a first-person perspective to mimic developer reasoning and be guided by SDLC principles with iterative refinement.
The prompt format for rewriting the reasoning steps is as follows, with detailed prompts provided in our replication package due to page limitation:
\begin{quote}
    \# \textbf{Overall Instruction}: Given the following coding task, the ground-truth code solution, and the reference approach description, generate the reasoning steps with the first-person perspective. The steps should follow the full development lifecycle: understanding the problem, planning the solution, implementing the code, testing thoroughly, and optimizing if necessary.
    
    \# \{\textbf{Instructions on SDLC Principles: ...}\}
    
    \# \{\textbf{Instructions on Generating Adaptive Iterative Refinements: ...}\}
    
    \# \{\textbf{Coding Task Description: ...}\}
    
    \# \{\textbf{Ground-Truth Code Solution: ...}\}
    
    \# \{\textbf{Augmented Discussion: ...}\}
    
    \end{quote}

Furthermore, in \dataset{}, we dynamically adjust the loop length of code refinement based on task complexity, assigning fewer refinement cycles to easier problems and more to harder ones.
This aligns with the common practice in competitive programming platforms, where problems are naturally categorized into varying difficulty levels.
Specifically, easy problems receive no refinement cycles, as they typically require straightforward implementations. Medium-level problems undergo one refinement cycle to capture minor optimizations and corrections, while hard problems are assigned two cycles to reflect the iterative debugging and improvement process.
For platforms without native task-difficulty labels, LLM-as-a-Judge~\cite{gu2024llmasjudge} can serve as a proxy, leveraging LLMs to assess and categorize task difficulty.
It is important to note that after fine-tuning with \dataset{}, LLMs do not require task-difficulty labels during inference. Instead, having learned reasoning steps adapted to varying complexities during training, the models inherently adjust their reasoning depth and length when encountering new, unlabeled problems.

\subsection{Proof-of-Concept Implementation}
\subsubsection{Dataset Creation}
Following the method of \dataset{}, we collect the reasoning dataset from the LeetCode platform with a focus on Python as a proof of concept.
We use DeepSeek V3 to conduct the completion and perturbation steps and QWQ~\cite{bai2023qwen} to rewrite and enrich augmented discussions.
We use two different LLMs due to our limited budgets for API services.
Specifically, we collect 2,105 LeetCode problems released before January 1, 2024, and filter 12,444 discussion posts according to the criteria detailed in \Cref{sec:svrc}. LeetCode problems released after this date are reserved for evaluation.
As a result, we obtain \dataset{}$_{LC}$, consisting of 12,444 data samples. Each sample contains 5,546 tokens on average, including the task description, reasoning chain, and corresponding code solution, tokenized using the OpenAI tokenizer.

\subsubsection{Dataset Quality Assessment}
We evaluate the quality of perturbed code solutions and reasoning chains on a representative subset of 377 samples from \dataset{}$_{LC}$, ensuring a statistically representative sample size with a 95\% confidence level and a 5\% margin of error.
Given that each sample contains thousands of tokens on average and encompasses intricate technical details that demand high-level domain expertise in coding algorithms, data structures, and programming languages, manual assessment is particularly labor-intensive.
To alleviate the burden of human annotation, we implement a hybrid evaluation strategy: we conduct a human evaluation to assess the correctness of code perturbations and the fluency of reasoning chains, complemented by an LLM-as-a-Judge~\cite{gu2024llmasjudge} assessment to evaluate the correctness of reasoning chains and their alignment with the corresponding code solutions.

\vspace{1mm}
\noindent\textbf{Human Evaluation.} 
We recruit two external annotators with at least two years of experience in Python development.
They independently assess the functional equivalence between the perturbed and original code with a binary classification label and rate each reasoning chain based on fluency  (i.e., clarity, completeness, structured presentation, and absence of unexpected characters), using a 0-5 scale, where higher scores indicate greater fluency.

\vspace{1mm}
\noindent \textbf{LLM-as-Judge Evaluation.}
As a complementary assessment, we employ the LLM-as-Judge approach~\cite{gu2024llmasjudge}, leveraging a state-of-the-art (SOTA) LLM to evaluate key data quality attributes. Given its superior performance and comparability to human annotators\cite{zheng2023judging}, GPT-4o is used to assess the logical correctness of reasoning chains and their alignment with the corresponding code solutions, assigning scores on a 0-5 scale, where higher scores indicate stronger correctness and coherence.

\begin{table}[t!]
    \centering
    \caption{Results of reasoning quality assessment using human and LLM-based evaluation.}
        \resizebox{0.41\textwidth}{!}{ % Resizing to fit text width
    \begin{tabular}{l c c}
        \toprule
        \textbf{Evaluation Method} & \textbf{Metric} & \textbf{Score (0-5)} \\
        \midrule
        \multirow{1}{*}{\textbf{Human Evaluation}} & Fluency & \textbf{5.00} \\
        \midrule
        \multirow{2}{*}{\textbf{LLM-as-Judge (GPT-4o)}} & Correctness & \textbf{4.89} \\
        & Alignment & \textbf{4.93} \\
        \bottomrule
    \end{tabular}}
    \label{tab:reasoning_quality}
    \vspace{-4mm}
\end{table}

\vspace{1mm}
\noindent \textbf{Results.}
For human evaluation, the fluency of reasoning chains was rated with an average score of 5.0, indicating high readability. 
Additionally, all perturbed code solutions remain functionally identical to their original versions. However, in two out of 377 cases, the generated textual format requirements do not correctly align with the perturbed code. In both instances, the instructions do not explicitly specify generating code in a particular format but rather require transforming the code from one version (the original) to another (the perturbed version).
We resolve this issue by scanning \dataset{}$_{LC}$ using keyword matching (e.g., convert, transform, rewrite, change), identifying 43 cases, which are then manually corrected.

% cases wrongly rephrase 
% the original code structure.
For LLM-based evaluation, GPT-4o assessed logical correctness with an average score of 4.89/5 and alignment with the final solution at 4.93/5, demonstrating that the reasoning chains are largely accurate and well-aligned with corresponding code solutions.

\subsubsection{Model Training}

With the dataset, we fine-tune target LLMs using supervised fine-tuning (SFT) with a next-token prediction objective. 
We refer to the LLM after fine-tuning as \tool{}.
This process enables the model to learn structured reasoning patterns before generating executable code, improving its problem-solving performance.

Given a training dataset \(\mathcal{D} = \{(x_i, y_i^+)\}_{i=1}^{n}\), where \( x_i \) represents the input prompt, \( y_i^+ \) is the corresponding target output, which consists of the structured reasoning chain followed by the code solution, the model is trained to maximize the likelihood of producing \( y_i^+ \) given \( x_i \), parameterized by \(\theta\). The supervised fine-tuning loss is defined as:
\begin{equation}
    \mathcal{L}_{SFT} = -\frac{1}{n} \sum_{i=1}^{n} \log P(x_i, y_i^+ | \theta),
\end{equation}
where \( P(x_i, y_i^+ | \theta) \) represents the model's predicted probability of generating the correct reasoning and code given the input \( x_i \). 

\subsubsection{Model Inference}
\label{sec:inference_prompting}
During inference, \tool{} follows a prompting strategy aligned with its fine-tuning process. 
Specifically, we explicitly ask LLM to generate reasoning steps guided by the SDLC principle with iterative refinement in the prompt.
The inference prompt follows the format shown below, with detailed prompts provided in our replication package.
\begin{quote}
    Your role as a coding assistant is to provide thorough, well-reasoned, and precise solutions to coding questions by following a systematic long-reasoning process. Please think step-by-step and carefully before arriving at the final answer. Use the following step-by-step workflow:

    \# \{\textbf{Instructions on SDLC Principles: ...}\}

    \# \{\textbf{Coding Task Description: ...}\}
\end{quote}

\section{Experiments Setting}

\subsection{Evaluation Dataset}
\label{sec:exp_setup_dataset}
Several widely-used benchmarks exist for evaluating code generation performance, such as HumanEval~\cite{chen2021humaneval}, MBPP~\cite{austin2021mbpp}, APPS~\cite{hendrycks2021apps}, CodeContests~\cite{li2022codecontests}, and LiveCodeBench~\cite{jain2024livecodebench}. These benchmarks typically provide test cases to facilitate execution-based evaluation. However, recent studies indicate that early benchmarks—including HumanEval, MBPP, APPS, and CodeContests—have likely been included in the training data of various LLMs, raising significant concerns regarding data contamination~\cite{jain2024livecodebench}, where evaluation datasets may have been exposed to LLMs during pretraining or fine-tuning.

To avoid this contamination risk, we use LiveCodeBench as the main evaluation dataset.
LiveCodeBench continuously crawls new code problems from diverse platforms to avoid potential data contamination.
% LiveCodeBench is a holistic and continuously updated benchmark designed to evaluate the code generation capacity of LLMs. Unlike static benchmarks, LiveCodeBench regularly incorporates new problems collected from weekly contests on platforms such as LeetCode, AtCoder, and Codeforces. 
% Each problem is tagged with its release date, ensuring evaluations are conducted on unseen problems, mitigating potential data contamination. 
The latest version of LiveCodeBench contains 713 coding problems from LeetCode, AtCoder, and Codeforces from May 2023 to September 2024.
Since \dataset{}$_{LC}$ is trained exclusively on LeetCode problems released before 2024, we select a subset of LiveCodeBench for evaluation.
This consists of all LeetCode problems released after January 1, 2024, as well as all AtCoder and Codeforces problems.
We refer to this subset of problems (with 562 coding problems) as LiveCodeBench in the following sections.
\subsection{Fine-tuning Settings}
We release \tool{}, which is fine-tuned with SVRC$_{LC}$, and use Qwen-2.5-32B-Instruct as the initial checkpoint. 
\tool{} is trained with full parameter updates for three epochs on a server with eight NVIDIA H100 GPUs. 
We follow the same fine-tuning settings as the default settings in LLaMA-Factory~\cite{llamafactory}, a commonly used LLM fine-tuning framework.
Specifically, the learning rate of \tool{} is set to 5e-5 with a cosine decay schedule, the warm-up ratio is set to 0.01, and gradient accumulation steps are set to 16.
Additionally, we analyze how different base models affect the performance of \tool{} in \Cref{sec: impactonsize}.

\subsection{Evaluation Baselines}
% We consider a set of LLMs as baselines. 
We evaluate \tool{} against a diverse set of baselines to assess its effectiveness in solving challenging code problems.

\vspace{1mm}
\noindent\textbf{SOTA Code LLMs}. We select open-source models with comparable parameter sizes to \tool{} and commercial closed-source LLMs for a comprehensive comparison.
Specifically, we include DeepSeekCoder-32B-Instruct~\cite{guo2024deepseek}, Qwen-2.5-32B-Instruct-Coder~\cite{hui2024qwen25coder} and CodeLlama-34B-Instruct~\cite{roziere2023codellama}, all of which are state-of-the-art open-source models known for their strong performance in code generation tasks.
Among those models, we highlight Qwen-2.5-32B-Instruct-Coder as it shares the same base model as \tool{} and is further fine-tuned on a trillion token-scale code corpus.
Furthermore, we report results on commercial models, including GPT-4o-mini-2024-07-18, Gemini-Pro-1.5, GPT-4o-2024-08-06, and GPT-o1-2024-12-07 to benchmark \tool{}'s performance against leading commercial LLMs on challenging code generation.

\vspace{1mm}
\noindent\textbf{Chain-of-Thought (CoT) based Approach.} We construct two CoT baselines for all open-source LLMs.
The first baseline follows the standard unsupervised CoT approach~\cite{wei2022chain}, asking LLMs to think step-by-step before producing the final code.
The second baseline is \tool{} style prompting as detailed in Section~\ref{sec:inference_prompting}, where the inference prompt explicitly instructs the LLM to generate SDLC-style reasoning steps before producing the final code.

\vspace{1mm}
\noindent \textbf{Reasoning-augmented LLM.} Finally, we establish LLMs with enhanced reasoning capabilities as baselines. For open-source models, we select S1, released by Stanford~\cite{muennighoff2025s1}, which has been fine-tuned on challenging questions from science and math datasets, along with corresponding reasoning traces and solutions synthesized via the Google Gemini Flash Thinking API~\cite{geminiflash}.
We highlight that S1 and \tool{} share the same base model, providing a direct basis for comparison.
For closed-source models, we select GPT-o1-preview, the SOTA reasoning-augmented LLMs to date.

\subsection{Evaluation Metrics}
Following standard practice, we report the pass@1 as the main metric. Pass@1 is the percentage of code solutions that pass all the given test cases at their first attempt.
Pass@1 ranges from 0 to 1, with higher values indicating better performance: 
\begin{equation} 
    \text{pass@1} = \frac{\text{Number of solutions that passed test cases}}{\text{Total number of coding problems}} 
\end{equation}

\section{Experiments Results}
\begin{table*}[ht]

    \caption{Performance of SOTA LLMs on LiveCodeBench. Easy, Medium, and Hard are the task difficulty labels of the LiveCodeBench dataset. We emphasize \tool{}'s performance gains over its base model using standard prompting in red, while the corresponding performance decline is shown in green.}    
    \begin{tabular}{ccccccc}
    \toprule              
    \multirow{2}{*}{Model}  &     \multirow{2}{*}{Prompting}   & \multirow{2}{*}{Size}         & Easy      & Medium     & Hard  & Overall   \\ 
       &    &           & pass@1    & pass@1    & pass@1  & pass@1     \\ \midrule
        \multicolumn{7}{c}{Open-source LLMs} \\ \hline
    CodeLlama-Instruct & Standard & 34B      & 0.290     & 0.028    & 0   & 0.078     \\ 
    CodeLlama-Instruct & CoT & 34B      & 0.298     & 0.028    & 0   & 0.106     \\ 
    CodeLlama-Instruct & \tool{} Style & 34B      &  0.298     & 0.035    & 0   & 0.109     \\ 

    s1-32B & Standard & 32B      & 0.306     & 0.098    & 0   & 0.134     \\ 
    s1-32B & CoT & 32B      & 0.315     & 0.112    & 0   & 0.142     \\ 
    s1-32B & \tool{} Style & 32B      &  0.315     & 0.105    & 0   & 0.141     \\ 
    
    DeepSeekCoder & Standard & 32B      & 0.589     & 0.126    & 0   & 0.235     \\ 
    DeepSeekCoder & CoT & 32B      & 0.565     & 0.133    & 0   & 0.233     \\ 

    DeepSeekCoder & \tool{} Style & 32B      & 0.601     & 0.133    & 0   & 0.243     \\ 
    Qwen2.5-Instruct       & Standard & 32B     & 0.847     & 0.343    & 0   & 0.398     \\ 
    Qwen2.5-Instruct       & CoT & 32B     & 0.823     & 0.371    & 0   & 0.401     \\ 
    Qwen2.5-Instruct       & \tool{} Style  & 32B     & 0.847     & 0.371    & 0   & 0.408     \\ 
        
    CodeThinker   & \tool{} Style    & 32B     & 0.831 \textcolor{red}{ (↓1.89\%)}     & \textbf{0.490}\textcolor{DarkGreen}{ (↑42.86\%)}    & 0   & \textbf{0.447}\textcolor{DarkGreen}{ (↑12.31\%)}       \\ \hline
    \multicolumn{7}{c}{Closed-source LLMs} \\ \hline
    GPT-4o-mini & Standard  & N/A     & 0.761     & 0.227    & 0.038   & 0.350    \\
    Gemini-Pro-1.5 & Standard      & N/A      & 0.838     & 0.245    & 0.042   & 0.384       \\
    GPT-4o & Standard &N/A       & 0.863     &0.283    & 0.056   & 0.410       \\
    GPT-o1 & Standard &N/A  & 1.000     & 0.667    & 0.435   & 0.672  \\    
    \bottomrule
    \end{tabular}    
    \label{tab:in_distribution_results}

\end{table*}

This section presents the evaluation results of \tool{} in comparison with state-of-the-art baselines. Specifically, our evaluation is designed to address the following research questions:

\begin{itemize}[leftmargin=*]
    \item RQ1: How does \tool{} compare to state-of-the-art LLMs in generating challenging code?    
    \item RQ2: What is the performance of \tool{} on challenging code problems out of the training distribution?
    \item RQ3: What is the impact of each component in \dataset{}? 

\end{itemize}
\begin{table*}[!t]
    \begin{threeparttable}

    \caption{Comparison between Qwen2.5-Coder and \tool{}.}
    \begin{tabular}{cccccc}
    \toprule              
    \multirow{2}{*}{Model}  &     \multirow{2}{*}{Fine-tune Dataset Scale}  & Easy      & Medium     & Hard  & Overall   \\ 
       &             & pass@1    & pass@1    & pass@1  & pass@1     \\ \midrule
       
    Qwen2.5-Coder & 5.5 trillion tokens      & 0.871     & 0.364    & 0   & 0.413     \\     
    CodeThinker   & 69 million tokens (1/80000)         & 0.831 \textcolor{red}{ (↓4.59\%)}     & \textbf{0.490}\textcolor{DarkGreen}{ (↑34.62\%)}    & 0   & \textbf{0.447}\textcolor{DarkGreen}{ (↑8.23\%)}       \\ \bottomrule
    \end{tabular}
    \label{tab:qwen2.5-coder_results} 
\end{threeparttable}

\end{table*}

\subsection{RQ1: Comparison with Other LLMs}

In this RQ, we compare the performance of \tool{} with state-of-the-art LLMs on LiveCodeBench.
% \subsubsection{Overall Performance Analysis}
Table~\ref{tab:in_distribution_results} presents the performance of selected LLMs across task difficulty in terms of pass@1.\footnote{We refer to the leaderboard of LiveCodeBench for the results of GPT-4o-mini and GPT-4o, the cut-off time of GPT-4o-mini is 07/08/2024, and GPT-4o is 06/08/2024.} We highlight the improvements \tool{} achieves over its base models in red and any performance regressions in green.

We observe that \tool{} achieves an overall pass@1 of 0.447, marking a 12.31\% improvement over its base model, i.e., Qwen2.5-Instruct-32B with standard prompting.
Besides, \tool{} also outperforms GPT-4o-mini, Gemini-Pro-1.5, and GPT-4o on all coding questions by 27.71\%, 16.41, and 9.02\% in terms of pass@1, respectively.
Notably, \tool{} significantly improves performance on solving medium-difficulty coding tasks, achieving performance improvement 
over its base model, GPT-4o-mini, and GPT-4o by 42.86\%, 73.14\%, and 115.86\% in terms of pass@1, respectively.
This substantial boost highlights \tool{}'s effectiveness in tackling more complex coding tasks that require structured reasoning and multi-step planning.

\tool{} also demonstrates on-par performance with its base model on easy tasks, with only a 1.89\% drop in pass@1, which is not statistically significant (p > 0.05). We attribute this minor decline to the nature of easy tasks, whose solutions do not require extensive reasoning. Therefore, the structured reasoning approach may introduce unnecessary verbosity, slightly reducing accuracy.

% \subsubsection{Effectiveness of \tool{}-Style Prompting}
We also notice that the prompting strategy {\em alone} has minimal impact on efficacy. 
Specifically, \tool{}-style prompting led to only minor improvements (less than 5\% in pass@1) for all LLMs compared to their base models with standard prompting. In contrast, CoT prompting resulted in inconsistent performance fluctuations.
The performance gap between \tool{} and its base model when both use \tool{}-style prompting—with \tool{} achieving a 32.1\% improvement in pass@1 on medium-difficulty tasks—suggests that inference prompting alone is insufficient to fully develop LLM reasoning abilities. Instead, fine-tuning plays a crucial role in equipping LLMs with effective reasoning skills.

We emphasize the importance of domain-specific fine-tuning datasets. Comparing S1 and \tool{}, we observe that S1, which is fine-tuned on science and math datasets with reasoning chains, performs worse than its base model on code generation tasks. In contrast, \tool{}, which is fine-tuned on reasoning chains specifically tailored for code generation, achieves significant performance gains over the same base model. This highlights the necessity of domain-specific reasoning augmentation for optimizing LLM performance in code generation and value of \dataset{}.

We also observe that fine-tuning on a small-scale but high-quality dataset of reasoning chains leads to better code reasoning abilities in LLMs compared to fine-tuning on a large corpus of raw code without structured reasoning. 
Specifically, we compare \tool{} against Qwen2.5-Coder, which shares the same base model as \tool{} but is fine-tuned on a 5.5-trillion-token code corpus. In contrast, \dataset{}$_{LC}$ contains only 69 million tokens, making it approximately 1/80,000 the size of Qwen2.5-Coder's training corpus.
Table~\ref{tab:qwen2.5-coder_results} presents the performance comparison. Notably, \tool{} demonstrates a 34.62\% improvement in pass@1 over Qwen2.5-Coder on medium-difficulty tasks, reinforcing the effectiveness of structured reasoning in handling more complex coding challenges. 
However, on easy tasks, \tool{} experiences a 4.59\% decline in pass@1 compared to Qwen2.5-Coder. This performance trend aligns with our hypothesis that solving easy tasks primarily relies on syntax and semantic understanding of the programming language, which Qwen2.5-Coder effectively acquires through extensive pre-training on a trillion-scale code corpus. In contrast, medium-difficulty tasks benefit significantly from a structured reasoning process, which is explicitly modeled in \dataset{}$_{LC}$.

\begin{tcolorbox}[tile, size=fbox,left=4mm, right=2mm, boxrule=0pt, top=2mm, bottom=2mm,
    borderline west={2mm}{0pt}{blue!70!black}, colback=blue!3!white, 
    sharp corners=south]
     \textbf{Answer to RQ1:} 
    \tool{} achieves significant improvements on challenging code problems. Specifically, \tool{} outperforms its base model (Qwen2.5-Instruct-32B), Qwen2.5-Coder, GPT-4o, and Gemini-Pro-1.5 on medium-difficulty code problems by 42.86\%, 34.62\%, 73.14\%, and 115.86\% in terms of pass@1, respectively.

\end{tcolorbox}

\subsection{Generalization to Non-LeetCode Platforms}
In this RQ, we evaluate the generalization ability of \tool{}. Since \dataset{}$_{LC}$ is exclusively constructed from LeetCode discussions, testing on non-LeetCode platforms provides insight into how well \tool{} adapts to unseen and diverse problem formats beyond its training distribution. 
Notably, the code solution format among code contest platforms varies significantly. For instance, LeetCode requires class-based solutions, where inputs are received as method parameters, and outputs are returned from methods. Explicit input/output statements are not needed in the solution. On the other hand, AtCoder mandates function-based solutions, where inputs are read using \textsc{input()} and outputs are printed using \textsc{print()}, which are required to be explicitly stated in the solution.

Specifically, we extract all coding problems on non-LeetCode platforms from LiveCodeBench. 
In total, we collect 394 code problems from AtCoder and Codeforces.
We apply the default format requirements of LiveCodeBench in the inference for all coding problems and compare the performance of \tool{} against its base model, Qwen2.5-Instruct-32B.

\begin{figure}[!t]
    \centering
    \includegraphics[width=0.5\textwidth]{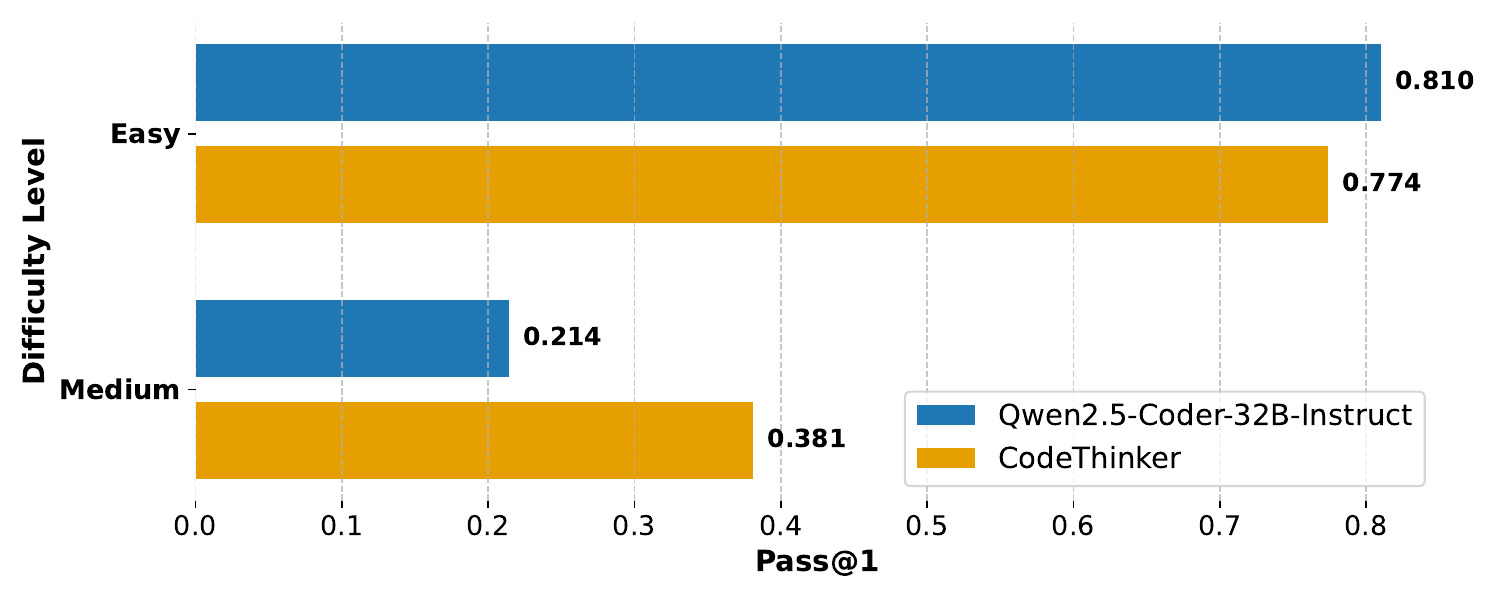}
    \caption{Comparison of CodeThinker and its base model on non-LeetCode coding problems (AtCoder and Codeforces), highlighting improved reasoning performance on medium-difficulty tasks.}
    \label{fig:out_distribution_results}
\end{figure}

\vspace{1mm}
\noindent \textbf{Results.} 
Our evaluation results is illustrated in Figure~\ref{fig:out_distribution_results}.
We omit the performance on hard-difficulty tasks as both models fail to solve any of them.
We observe that \tool{} demonstrates a strong generalization ability to non-LeetCode platforms and shows a similar trend as which is observed on full LiveCodeBench. \tool{} shows on-par performance with its base model on easy code problems, i.e., 4.44\% drop in terms of pass@1, while achieving significant improvements on challenging code problems, i.e., 78.01\% improvement in terms of pass@1 on medium-difficulty tasks.
This suggests that the structured reasoning capabilities of \tool{} still provide a great advantage when tackling challenging code problems from non-LeetCode platforms. 
However, the slight performance drop on easy code problems indicates that the unnecessary reasoning chains generated by \tool{} for easy code problems may introduce additional cognitive load, leading to slight performance degradation.

\begin{tcolorbox}[tile, size=fbox, left=4mm, right=2mm, boxrule=0pt, top=2mm, bottom=2mm,
    borderline west={2mm}{0pt}{blue!70!black}, colback=blue!3!white, 
    sharp corners=south]
     \textbf{Answer to RQ2:} 
    \tool{} shows strong generalization ability to problems from non-LeetCode platforms, achieving significant improvements on challenging code problems, i.e., 78.01\% improvement in pass@1 on medium-difficulty tasks over its base model.
\end{tcolorbox}

\subsection{RQ3: Ablation Study}
% \subsubsection{Effectiveness of each stage of \dataset{}}
\label{sec:ablation1}
In this research question, we aim to quantify the impact of each stage of crafting \dataset{} on the performance of \tool{}. Specifically, we conduct an ablation study by fine-tuning Qwen2.5-Instruct-32B using different stages of \dataset{} and evaluating the resulting models on LiveCodeBench under \tool{}-style prompting.

\begin{figure}[!t]
    \centering
    \includegraphics[width=0.46\textwidth]{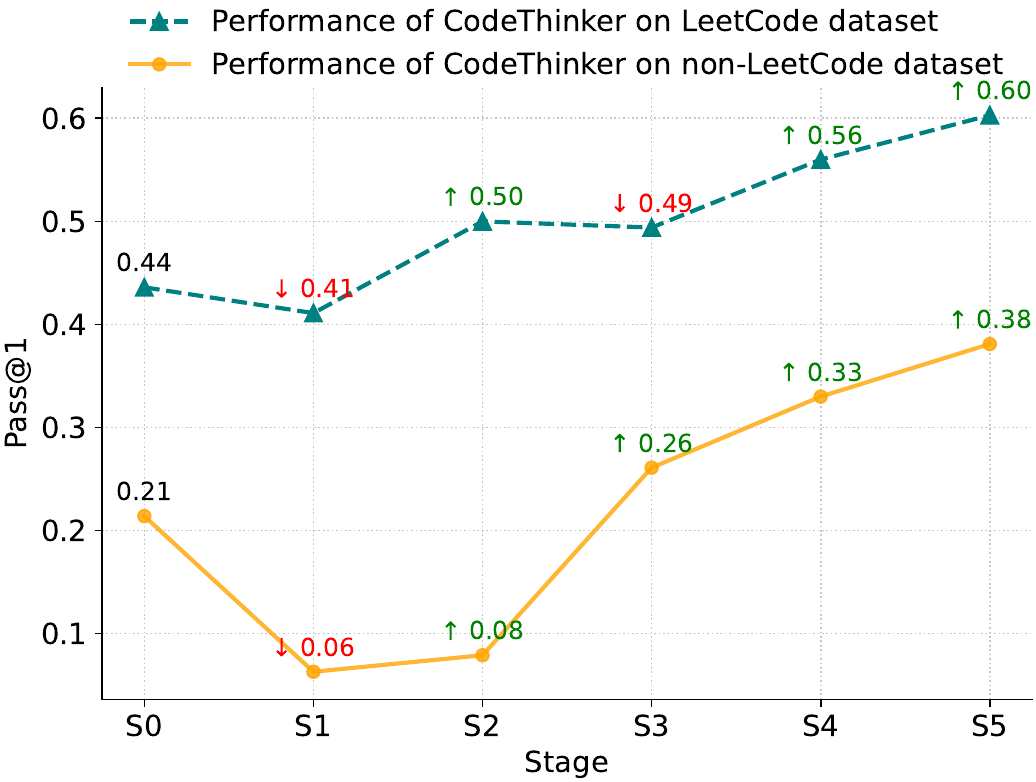}
    \caption{Performance of \tool{} on different stages of \dataset{}. The performance drop from S0 to S1 further suggests that LLMs struggle to learn reasoning patterns solely from raw community data, which motivates our design of \dataset{}.}
    \label{fig:ablation_study}
\end{figure}

We define the following progressive stages in constructing \dataset{}:
\begin{itemize}[leftmargin=*]
    \item \textbf{S0 Initial Checkpoint:} Apply Qwen2.5-Instruct-32B without fine-tuning.
    \item \textbf{S1 Raw LeetCode discussions}: Utilize raw LeetCode discussions for fine-tuning.
    \item \textbf{S2 Discussion Completion}: Utilized discussions in which missing sections are completed for fine-tuning
    \item \textbf{S3 Fromat Perturbation}: Discussions that are completed and pertured are used for fine-tuning.
    \item \textbf{S4 Reasoning chains aligned with SDLC principles without iterative refinement}: 
    Fine-tuning with reasoning chains follows SDLC principles but without introducing iterative refinement cycles.
    \item \textbf{S5 Full SVRC$_{LC}$}: Use full \dataset{}$_{LC}$ for fine-tuning.
\end{itemize}

We follow the dataset split setting in RQ2, categorizing LiveCodeBench into LeetCode and Non-LeetCode datasets. The LeetCode dataset consists of problems released after January 1, 2024.

\vspace{1mm}
\noindent \textbf{Results.} 
We observe that fine-tuning with any stage of \dataset{} has limited impact on easy and hard-level code problems, with a performance oscillation of less than 10\% on easy problems and 0 successful attempts on hard ones. Therefore, we present the pass@1 performance of \tool{} for medium-difficulty code problems across different stages of \dataset{}, as illustrated in Figure~\ref{fig:ablation_study}. The results reveal several key insights. 

1. \textbf{Fine-tuning with raw LeetCode discussions (S1) results in a performance drop compared to the base model (S0).} This is expected as many LeetCode discussions are incomplete and unstructured, which introduces noise into the fine-tuning process. 

2. \textbf{Enhancing discussions with completed insights (S2) improves the performance on LeetCode problems.} The model demonstrates a 21.96\% increase in pass@1 from S1 to S2.
However, this improvement is primarily observed in the LeetCode dataset, with minimal impact on non-LeetCode performance due to poor data diversity. We manually inspect the code generated for non-LeetCode data. 
We observe that the generated codes are incorrectly structured as a Python class even though the task requirements ask for code to be provided as a function. 
This is caused by how all ground-truth solutions of S2 are written as Python classes.

3. \textbf{Applying perturbation techniques (S3) improves generalization to non-LeetCode problems.}
The pass@1 score on the non-LeetCode dataset improves from 0.08 to 0.26, confirming the importance of perturbation for enhancing model robustness. However, this step slightly reduces performance on the LeetCode dataset, suggesting a trade-off between generalization and specialization.

4. \textbf{Aligning reasoning chains with SDLC principles (S4) leads to substantial performance gains.} The transition from S3 to S4 results in a 22.44\% increase in pass@1 on the LeetCode dataset, while further refinement from S4 to S5 yields a 46.15\% improvement on the non-LeetCode dataset. This underscores the effectiveness of structuring reasoning processes using SDLC guidelines.

\begin{tcolorbox}[tile, size=fbox, left=4mm, right=2mm, boxrule=0pt, top=2mm, bottom=2mm,
    borderline west={2mm}{0pt}{blue!70!black}, colback=blue!3!white, 
    sharp corners=south]
     \textbf{Answer to RQ3:} 
 The ablation study highlights that each stage of \dataset{} crafting contributes to the final performance of \tool{}. 
 Reasoning chain rewrite and iterative refinements lead to improvements of 22\% and 46\%, respectively.  
 \end{tcolorbox}

\section{Discussion}
\subsection{The impact of LLM size}
\label{sec: impactonsize}
We investigate how model size and code capacity influence fine-tuning effectiveness for code reasoning.
Specifically, we evaluate two smaller variants of the Qwen2.5 family—Qwen2.5-7B-Instruct and Qwen2.5-14B-Instruct—as well as DeepSeekCoder-32B-Instruct, which shares the same parameter size as \tool{}'s base model but demonstrated inferior performance in RQ1.
Following the methodology outlined in Section~\ref{sec:ablation1}, we fine-tune these models using the full \dataset{} and compare their performance on medium-level code problems of LiveCodeBench against their base versions.

\begin{table}[!t]
\caption{Performance of LLMs of varying sizes on LiveCodeBench. 
Base version refers to the initial checkpoint; \dataset{} version refers to the LLM after fine-tuning with \dataset{}.}
\resizebox{0.49\textwidth}{!}{

\begin{tabular}{llll}
\toprule
    & Version & LeetCode & Non-LeetCode \\ \midrule
\multirow{2}{*}{Qwen2.5-7B} & base        & 0.208                & 0.101                    \\ 
& \dataset{}        &  0.321 (\textcolor{DarkGreen}{↑54.32\%})  & 0.026 (\textcolor{red}{↓74.26\%})\\ \cmidrule(lr){2-4}
\multirow{2}{*}{Qwen2.5-14B} & base        & 0.308                & 0.161                    \\ 
& \dataset{}        &  0.551 (\textcolor{DarkGreen}{↑78.89\%})  & 0.114 (\textcolor{red}{↓41.23\%})\\ \cmidrule(lr){2-4}
\multirow{2}{*}{DeepSeekCoder-32B} & base        & 0.115                & 0.147                    \\ 
& \dataset{}        &  0.291 (\textcolor{DarkGreen}{↑153.04\%})  & 0.161 (\textcolor{DarkGreen}{↑9.52\%})\\ \cmidrule(lr){2-4}
\multirow{2}{*}{Qwen2.5-32B} & base        & 0.436                & 0.214                    \\ 
& \dataset{}        &  0.603 (\textcolor{DarkGreen}{↑38.30\%})  & 0.381 (\textcolor{DarkGreen}{↑78.04\%})\\ \bottomrule

\end{tabular}
}
\label{table:llmsize}
\end{table}

\vspace{1mm}
\noindent \textbf{Results.} 
% We summarize two key findings here.
The results presented in Table~\ref{table:llmsize} reveal two primary trends regarding the influence of model size on fine-tuning efficacy.
First, the relative improvement of LeetCode performance from fine-tuning is inversely correlated with the base model’s capacity.
DeepSeekCoder-32B, which exhibited the lowest initial performance among the selected models, achieved the highest relative improvement after fine-tuning, with a 153.04\% increase in pass@1 on the LeetCode dataset. 
Conversely, Qwen2.5-32B, which already demonstrated strong base performance, saw a more moderate improvement of 38.30\%. This suggests that models with weaker initial reasoning capabilities derive greater benefit from fine-tuning with structured reasoning data. 
However, the best initial model still leads to the best absolute performance. 

Second, the impact of fine-tuning on non-LeetCode generalization varies significantly based on model size.
The efficacy of Qwen2.5-7B and Qwen2.5-14B on the non-LeetCode dataset deteriorates after fine-tuning. In contrast, DeepSeekCoder-32B and Qwen2.5-32B demonstrate improved generalization, with their pass@1 increasing by 9.52\% and 78.04\%, respectively. These results indicate that smaller models are more prone to overfitting, whereas larger models exhibit stronger generalization capacity.
One potential explanation for this discrepancy lies in the nature of perturbations applied to \dataset{}. While the dataset introduces variation in input, output, and structure of code solutions, it does not alter the problem descriptions. Consequently, smaller models, which may rely more on textual memorization, may struggle with code problems written in different styles.

% \subsection{Reasoning Quality Assessment}
\label{sec:reasonquality}

\subsection{Threat to Validity}
Threats to external validity pertain to the generalizability of \dataset{}. To mitigate this threat, we fine-tune LLMs of different sizes and architectures, demonstrating that \dataset{} enhances LLMs' reasoning capabilities. Moreover, as \dataset{}$_{LC}$ is sourced from LeetCode, we evaluate \tool{} on coding problems from platforms other than LeetCode, i.e., AtCoder and CodeForces, and confirm the generalizability of \tool{}. Threats to internal validity primarily stem from potential biases introduced during the data collection.
In particular, the automated extraction and structuring of reasoning chains from community discussions might introduce unintended inaccuracies or inconsistencies. To address this, we employ heuristic filtering rules to retain high-quality discussions and carefully designed prompts grounded in software engineering best practices to ensure structured, valid reasoning chains. Furthermore, we validate the quality of the resulting dataset through manual inspections and quantitative assessments, confirming the consistency and correctness of the extracted reasoning steps.

\section{Conclusion and Future Work}
In this work, we introduce SVRC, a novel framework that systematically extracts, structures, and enriches reasoning chains from software engineering community discussions to improve LLMs' code reasoning abilities. By integrating Software Development Life Cycle (SDLC) principles, SVRC enables LLMs to engage in iterative problem-solving, thereby enhancing their ability to tackle complex, multi-step coding tasks.
Building upon SVRC, we proposed CodeThinker, a reasoning-augmented LLM fine-tuned with SVRC, which outperforms state-of-the-art models—including GPT-4o, Gemini-Pro-1.5, and Qwen2.5-32B-Instruct—on LiveCodeBench's code generation tasks.
Our ablation study further confirms that each stage of \dataset{} contributes to the final performance of \tool{}.

Moreover, we aim to explore further opportunities to refine our reasoning chains by investigating alternative software development principles and integrating domain-specific expert knowledge, enabling even more generalizable and robust reasoning-augmented code generation frameworks.

% \section{Data Availability}
% \label{sec:replication}
% The implementation of \dataset{}, along with our SVRC$_{LC}$ dataset and \tool{} model, have been anonymized and made available at
% \begin{center}
%     \url{https://anonymous.4open.science/r/SVRC-ICSE/}
% \end{center}

\bibliographystyle{plain}
% \bibliography{bibliography}
\bibliography{bib}

\end{document}